\documentclass[a4paper,11pt]{article}

\usepackage{graphicx}
\usepackage{caption}
\usepackage{subcaption}
\usepackage{slashed,nicefrac}

\usepackage{mathtools}

\usepackage{fullpage}
\usepackage[T1]{fontenc}

\usepackage{amssymb,amsmath,cite}

\usepackage{xcolor,bbold,comment}

\usepackage{amsthm}

\newcommand{\be}{\begin{eqnarray}}
\newcommand{\ee}{\end{eqnarray}}
\allowdisplaybreaks

\usepackage[utf8]{inputenc}
\usepackage[english]{babel}

\numberwithin{equation}{section}
\begin{document}

\begin{titlepage}
	\thispagestyle{empty}
	\begin{flushright}
		
	%	\hfill{DFPD-2020/TH/XX} 
	\end{flushright}

	\vspace{35pt}
	
	\begin{center}
	    { \Large{\bf On Global Symmetries and Fayet--Iliopoulos Terms  }} 
		
		\vspace{50pt}
		
		{Fotis~Farakos, Alex~Kehagias and Nikolaos~Liatsos} 
		
		\vspace{25pt}

        {\it   Physics Division, National Technical University of Athens \\
        15780 Zografou Campus, Athens, Greece}
		
		\vspace{15pt}

		\vspace{40pt}
		
		{ABSTRACT} 
	\end{center}

We revisit the genuine Fayet--Iliopoulos terms of 4D N=1 supergravity. 
Such terms are commonly believed to preserve a global symmetry, 
and   therefore  they are in conflict with the principles of quantum gravity. 
However, we find that generically there do exist supersymmetric terms that break explicitly the specific global symmetry, while preserving gauge invariance. 
We illustrate this, by  providing a series of examples, including superpotentials and superspace higher order terms, 
along with a general prescription for their construction.

\vspace{10pt}

\bigskip

\end{titlepage}

%\newpage
%\tableofcontents

\newpage

\section{Introduction}

The current research direction in the formal four-dimensional supergravity literature is often focused on the search for ultra-violet (UV) restrictions that are imposed on the low energy supergravity effective theories \cite{Cribiori:2023gcy}. 
These restrictions, having a UV origin, are usually based on swampland criteria \cite{Vafa:2005ui,Agmon:2022thq,VanRiet:2023pnx}. 
This endeavor lies on the pattern that shows the effectiveness of the swampland criteria getting amplified within a supergravity theory. 
This happens due to supersymmetry, which connects the various terms with each other, and 
therefore a restriction on a gauge coupling may turn out to also restrict the Yukawa couplings or the vacuum energy, 
and so on \cite{Wess:1992cp,Kuzenko:1998tsq,Freedman:2012zz,DallAgata:2021uvl}. 
This is a supergravity effect that does not in principle occur in an arbitrary gravitational effective field theory (EFT).

The typical swampland criteria are basic restrictions on the gauge couplings, 
the existence of global symmetries, 
the moduli space of an EFT, 
etc. 
Such simple restrictions on the validity of an EFT, 
as for example the weak gravity conjecture (WGC) \cite{Arkani-Hamed:2006emk,Harlow:2022ich}, 
may turn out to have far-reaching implications for an effective supergravity theory. 
Indeed, as was shown in \cite{Cribiori:2020wch}, with the use of the WGC one can eliminate simple de Sitter vacua of pure 4D N=1 supergravity when they are supported by a Fayet--Iliopoulos (FI) term \cite{Freedman:1976uk}. 
The extension to 4D N=2 de Sitter was performed and discussed in a series of works \cite{Cribiori:2020use,Montero:2021otb,DallAgata:2021nnr}, 
highlighting the relation also to other swampland criteria \cite{Cribiori:2021gbf,Castellano:2021yye}. 
Restrictions also on supersymmetric vacua of 4D/5D supergravity due to the WGC were analyzed in \cite{Cribiori:2022trc,Cribiori:2023ihv}. 
For a series of further works where one can see the interplay between the swampland program and the structure of supergravity 
leading to interesting results see e.g. \cite{Ferrara:2019tmu,Lanza:2019xxg,Palti:2020tsy,Lanza:2020qmt,Martucci:2022krl,Montero:2022ghl,Dalianis:2023ewd,Casas:2024ttx}.

From our previous discussion we see that it is crucial to have a good account of the precise restrictions that are actually imposed on supergravity theories due to the swampland. 
Here our focus will be on the presence of unwanted global symmetries. 
It is a common lore that a theory of quantum gravity should not have any global symmetries (discrete or continuous; see e.g. \cite{Banks:2010zn}) and one should test supergravity EFTs against such a restriction as well. 
Indeed, due to the structure of supergravity, an unwanted (possibly accidental) global symmetry may appear. 
Typically, one is not worried about such global symmetries when they show up in a low energy EFT because we assume that 
there will exist some higher order term, probably Planck-suppressed, that will eventually violate the global symmetry. 
If instead the consistency of a model is rooted in the preservation of the global symmetry 
at all orders, then such a model would not be consistent with a theory of quantum gravity.

Such a problematic global symmetry is discussed in \cite{Komargodski:2009pc}, 
and further aspects of it are analyzed in \cite{Dienes:2009td,Seiberg:2010qd,Distler:2010zg,Kuzenko:2009ym} (including supercurrent properties and charge quantization). 
This symmetry appears in the FI-gauged supergravity and will be of central interest in this work. 
The specific system in question has to do with the embedding of the FI term \cite{Fayet:1974jb} within supergravity and the analysis of its salient properties (for some details of the FI terms in supergravity specifically, see e.g. \cite{Barbieri:1982ac,Cremmer:1983yr,Binetruy:2004hh,VanProeyen:2004xt,Villadoro:2005yq,Catino:2011mu,Cribiori:2017laj}). 
The outstanding observation of \cite{Komargodski:2009pc} is that in systems with a genuine FI term (i.e. an FI term that is not generated due to some Stueckelberg effect), not  maintain a global symmetry at the two-derivative supergravity level, 
but moreover, 
 such symmetry is preserved in all higher order terms. 
The latter is problematic because it means that the gravity-coupled theory will always preserve the global symmetry and therefore cannot be consistently coupled to quantum gravity. 
It is important to note that this specific result holds for matter-coupled supergravity with charged fields \cite{Seiberg:2010qd}. 
Indeed, if one works with a pure supergravity coupled only to a vector multiplet and a FI term,  then there is no global symmetry at all, but that happens just because there is no matter sector present in the theory. 
In addition, 
the analysis of \cite{Komargodski:2009pc} focuses on the situation where the FI term survives in the rigid supersymmetry limit, 
which means that the FI parameter does not scale with $M_P$ and is a small perturbative coupling.

In this work, we will see that indeed a global symmetry can be conserved in a supergravity theory 
but it is by no means a mandatory requirement of supergravity per se. 
We will show the latter fact simply by writing down specific higher order terms that explicitly break such a global symmetry, 
while preserving the gauge invariance related to the FI term. 
We will also make sure that such terms have a consistent rigid supersymmetry limit and that they do not require the presence of a 
massive vector (thus making the FI term effectively non-genuine). 
As we will see, there are actually infinite terms that one can write down with such properties; 
therefore, even if a 2-derivative supergravity has a global symmetry, there is no reason for the full higher derivative supergravity EFT to respect it. 
As a result, there is no direct conflict of the genuine FI terms with the principles of quantum gravity, 
at least as far as global symmetries are concerned.\footnote{Indeed, as we discussed earlier, 
it is now known that FI gauged supergravity can be in conflict with the WGC \cite{Cribiori:2020wch} or with charge quantization \cite{Seiberg:2010qd,Distler:2010zg}.}

\section{FI terms in 4D N=1}

Following the conventions of \cite{Wess:1992cp}, 
we work with chiral superfields $\Phi^i$, which contain as physical fields the complex scalars $A^i$ and the Weyl spinors $\chi^i_\alpha$. 
The gauge sector is described at the superspace level by the real gauge superfield $V$, which in turn contains as physical fields  
the abelian gauge vector $A_m$, and the gaugino $\lambda_\alpha$, while its field strength is the chiral superfield 
$$W_\alpha(V) = -\frac14 (\overline{\cal D}^2 - 8 {\cal R}) {\cal D}_\alpha V.$$ 
The gauge superfield $V$ transforms under a gauge transformation as 
\be
V \to V + i \Lambda - i \overline \Lambda \,, 
\ee 
where $\Lambda$ is a chiral superfield.  
If the theory is invariant under  global transformations of  the chiral superfields  of the form
\be
\label{C0}
\Phi^i \to \Phi^i \times \exp[i Q_i c_0] \,, 
\ee
for some real constant $c_0$, 
then this symmetry can be gauged by setting 
\be
\Phi^i \to \Phi^i \times \exp[i \Lambda Q_i] \,. 
\ee
Then, for example, we have the invariant combination 
\be
\label{flat-gauged}
\overline \Phi^i e^{- Q_i V} \Phi^i \,. 
\ee
Notice that \eqref{C0} implies the well-known charge degeneracy among the component fields of a chiral multiplet.

The sigma-model geometry is captured by the K\"ahler potential $K$, 
which is a real function of $\Phi^i$ and $\overline \Phi^i$.
Depending on the isometries of the sigma-model target space and the way we gauge them, we introduce  the ``gauged'' K\"ahler potential $K+ \Gamma$, 
where $\Gamma$  depends also on $V$. 
One example of $K + \Gamma$ is given by the expression \eqref{flat-gauged} for a flat K\"ahler potential with a gauged abelian isometry. 
However, 
the gauged K\"ahler potential does not need to be fully invariant under the gauge transformation, 
instead it can be invariant up to a K\"ahler transformation, 
and this is the window that allows to introduce a Fayet--Iliopoulos (FI) term. 
In particular, a ``genuine'' FI term will have the form 
\be
K + \Gamma = \dots + \xi V \,, 
\ee
for a real constant $\xi$, with 
\be
\label{genFI}
\delta_\Lambda ( K + \Gamma ) = i \xi \Lambda - i \xi \overline \Lambda \,. 
\ee
In a globally supersymmetric theory, such a coupling does not lead to any complications; 
however, 
in supergravity, due to the fact that gauge invariance is maintained only due to a super-Weyl rescaling, the FI term leads to some intricacies that we will now analyze.

The standard matter-coupled two-derivative supergravity theory has the form 
\be
\begin{aligned}
\label{stand}
{\cal L}_\text{standard} =& - 3 M_P^2 \int d^4 \theta E \exp[ - (K + \Gamma) / 3 M_P^2 ] 
\\
& + \left[ \int d^2 \Theta 2 {\cal E} \left( \frac{1}{4g^2} W^2 
+ W(\Phi^i) \right) + c.c. \right]  . 
\end{aligned}
\ee
Due to the overall K\"ahler transformation \eqref{genFI} that we get from the FI term, 
it is necessary to perform an additional super-Weyl rescaling to maintain gauge invariance. 
This rescaling has the following effects \cite{Wess:1992cp,Kuzenko:1998tsq} 
\be
d^4 \theta \, E \rightarrow d^4 \theta \, E \, e^{2 \Sigma + 2 \overline \Sigma} 
\quad , \quad  
d^2 \Theta \,  2{\cal E} \rightarrow d^2 \Theta \,  2{\cal E} \, e^{6 \Sigma} \, , 
\ee
while $$W^2(V) \to e^{-6 \Sigma} W^2(V).$$ Therefore, we can easily verify that the  first two terms in the Lagrangian \eqref{stand} are gauge invariant if  
\be
\label{R-G}
\Sigma = \frac{i  \xi \Lambda}{6 M_P^2} \,.  
\ee
In addition, the superpotential has to change under the gauge transformation as  
\be
\label{W-FI} 
W(\Phi^i) \to W(\Phi^i) e^{-6 \Sigma} = W(\Phi^i) e^{- i  \xi \Lambda / M_P^2} \,, 
\ee
in order for the full Lagrangian \eqref{stand} to be gauge invariant. 
The transformation \eqref{W-FI} eventually enforces the appropriate R-symmetry charges of the various superfields together with the K\"ahler potential. 
Indeed, 
if we use the R-charges of the chiral superfields defined as
\be
\Phi^i \to e^{3 R_i \Sigma} \Phi^i \,, 
\ee
then we clearly need the sum of the R-charges of each individual term that appears in the superpotential to be equal to to $-2$.

Let us now see how the continuous global symmetry will enter the setup, 
leaving its breaking for the next section. 
Let us first of all assume that the starting theory has in any case a global symmetry given by \eqref{C0}, 
which is further gauged by the $V$ multiplet, 
which on top of that has a genuine FI term. 
Then, due to the identification \eqref{R-G}, a general gauge transformation will act on the chiral superfields as   
\be
\label{SuperfieldG}
\Phi^i \to \exp\left[i \Lambda \left(Q_i + \frac{\xi R_i}{2 M_P^2}\right)\right] \times \Phi^i \,. 
\ee
As a result, we see that the gauge charges of the chiral superfields are shifted to the new charges 
\be
\label{tildeQ}
\tilde Q_i = Q_i + \frac{\xi R_i}{2 M_P^2} \,. 
\ee
Actually, 
there is also a lift of the charge degeneracy within each chiral multiplet. 
Indeed, 
from the definition of the component fields through superspace derivatives we have 
that the physical fields are given by 
\be
  A^i =  \Phi^i |_{\theta=0} \ , \ \chi^i_\alpha = \frac{1}{\sqrt 2} {\cal D}_\alpha \Phi^i |_{\theta=0} \,. 
\ee
Now, taking into account that ${\cal D}_\alpha$ rotates under a super-Weyl transformation as 
\be
\label{SWD}
{\cal D}_\alpha (\text{Scalar}) \to e^{\Sigma - 2 \overline \Sigma} {\cal D}_\alpha (\text{Scalar}) \,, 
\ee
and identifying the parameter $\lambda(x)$ of the gauge transformation as 
\be
\Lambda(x,\Theta) =  \lambda(x) + \dots \,, 
\ee
where the dots represent contributions that we use to restrict the gauge multiplet to the Wess--Zumino gauge, 
we have that the charges of the physical component fields are 
\be
\label{ShiftB}
A^i \to \exp\left[ i \lambda \tilde Q_i \right]  A^i \,, 
\ee
and 
\be
\label{ShiftF}
\chi^i_\alpha \to \exp\left[ i \lambda \left(\tilde Q_i + \frac{\xi}{2 M_P^2} \right) \right] \chi^i_\alpha \,. 
\ee
For completeness, 
let us note that the gravitino and the gaugino rotate as 
$$\psi_m^\alpha \to \psi_m^\alpha e^{- \frac{i}{2 M_P^2} \lambda \xi }, \qquad  \mbox{and}\qquad \lambda_\alpha \to \lambda_\alpha e^{- \frac{i}{2 M_P^2} \lambda \xi},$$ respectively.  
Now we can see the global symmetry that was highlighted in \cite{Komargodski:2009pc,Seiberg:2010qd}. 
Indeed, 
the original theory has a global symmetry that is given by \eqref{C0} to start with, 
before the gauging is introduced. 
However, once we gauge the abelian symmetry \eqref{C0} and introduce a FI term, the charges are not $Q_i$ any more, 
but instead they become \eqref{ShiftB} for the scalars and  \eqref{ShiftF} for the fermions. 
The actual issue is not that the charges change per se, 
but that now the charge of each fermion is unavoidably different from the charge of its scalar superpartner, due to the excess $\frac{\xi}{2}$ originating from the super-Weyl rescaling \eqref{SWD}.

We therefore see that the rotations for $\xi=0$ and $\xi \ne 0$ are manifestly different, 
and the first one represents a global symmetry while the second case is the FI-gauged rotation. 
In \cite{Komargodski:2009pc} it is asserted that in principle the global symmetry with $\xi=0$ is an 
intrinsic requirement for the consistency of the structure of the supergravity theory 
and that it goes hand-in-hand with the gauging. 
As a result, it will be respected by all higher order / higher derivative terms, 
and therefore it will ultimately lead to a quantum gravity theory with a global symmetry. 
Since the latter is claimed to belong to the swampland, 
it seems that the fate of FI-gauged supergravity is to be in the swampland as well. 
Notably, in \cite{Seiberg:2010qd} some cases where a global symmetry does not appear were discussed.  
However, 
that was focused on the situation where $\xi$ is quantized in terms of the Planck mass \cite{Distler:2010zg} (see also \cite{Dienes:2009td}).

The authors of \cite{Komargodski:2009pc} specifically focus on the setup where a formal global limit exists, 
such that 
\be
\label{global}
M_P \to \infty \ , \quad \xi = \text{finite} \,. 
\ee
Of course, in this case, one is faced with the issue of not having a compact abelian group; 
however, we set this issue aside until the discussion section and focus on the existence of a continuous global symmetry. 
Actually, what we will show now is that the global symmetry can be broken with a large variety of different terms while maintaining the gauge invariance. 
Therefore, the existence of the global symmetry is unrelated to the FI-gauged U(1) and there is no reason that such global symmetry is respected by the higher order string theory / quantum gravity corrections. 
In other words, 
the global symmetry could appear accidentally in the low energy theory but there is no reason related to the structure of supergravity that will forbid its breaking. 
It is actually exactly the fact that the global symmetry and the gauged one rotate the matter fields differently that allows 
to explicitly break one while keeping the other.

\section{Breaking the global symmetry}

As we have claimed, 
since the global symmetry is not needed for the consistency of the gauging, it should be possible to break it, 
and actually there are infinite kinds of terms that one can write down. 
Here we will only illustrate some simple examples of such terms. 
We want to stress right away that we are agnostic about the actual term that string theory (or some theory of quantum gravity in general) 
will provide in order for the global symmetry to be broken. 
Instead, 
the mere fact that terms that break the global symmetry while respecting the FI-gauged one exist proves the point 
that the two symmetries are independent as far as classical supergravity is concerned.

The first term that we can discuss is a contribution to the superpotential. 
Let us assume that we have at least one chiral superfield in our theory, say $\Phi^1$, 
and let us call it here simply $\Phi$ skipping the chiral-model index. 
Then, this superfield unavoidably also transforms as \eqref{SuperfieldG}, 
and we notice that under the FI-gauged abelian rotation we have 
\be
\left( \frac{\Phi}{M_P} \right)^{-\frac{\xi}{\tilde Q M_P^2}} \to 
e^{- i  \xi \Lambda / M_P^2} \times 
\left( \frac{\Phi}{M_P} \right)^{-\frac{\xi}{\tilde Q M_P^2} }  \,. 
\ee
Here we are implicitly assuming that $\tilde Q = Q + \frac{\xi R}{2 M_P^2}  \ne 0$; 
if instead $Q + \frac{\xi R}{2 M_P^2} = 0$, the situation is much easier, as we will see momentarily. 
From the transformation of the above term we see that we can have an extra term in the superpotential of the form 
\be
\label{W-break}
W_\text{break} = b_0 \left( \frac{\Phi}{M_P} \right)^{-\frac{\xi}{\tilde Q  M_P^2} } \,, 
\ee
which will preserve the gauge invariance because it transforms exactly as in \eqref{W-FI} (here $b_0$ is some constant with $[b_0]=3$). 
On the contrary, 
due to the presence of the $\Phi$ term, the global symmetry is explicitly broken. 
Note that if we take the formal global limit \eqref{global}, then this term becomes a constant (assuming that $b_0$ does not scale), 
which nevertheless is irrelevant in the formal global supersymmetry limit. 
If instead we have $Q + \frac{\xi R}{2 M_P^2} = 0$, then $\Phi$ is neutral under the gauged FI symmetry, while it is not neutral under the global rotation (we are still assuming $Q \ne 0$). 
In such a case, we can simply break the global symmetry with a gauge kinetic function, e.g. of the form 
\be
\int d^2 \Theta 2 {\cal E}  \left( \frac{1}{g^2} + \frac{\Phi}{M_P} \right) W^2 \,, 
\ee
or a superpotential of the form 
\be
W_\text{break} \sim  \frac{\Phi}{M_P} \times W(\Phi^i) \,, 
\ee
assuming that our theory has a non-trivial superpotential. 
Note that we are assuming that $Q \ne 0$ because if $Q = 0$ then there is no global symmetry to start with. 
All in all, 
what we have seen is that one can easily find superpotentials that break the global symmetry of \cite{Komargodski:2009pc} while 
maintaining the abelian FI-gauged one. 

One could however be dissatisfied with some properties of the $W_\text{break}$ contribution to the superpotential. 
Indeed, it may have some pathologies, which one would be tempted to extrapolate to the full theory. 
For example, 
it is not obvious that the superpotential and the full supergravity Lagrangian is well-defined in the limit 
\be
\label{Ato0}
\langle A \rangle  \to 0 \, , 
\ee
because of the way that $\Phi$ enters in \eqref{W-break}. 
We want to stress that being able to formally take the limit \eqref{Ato0} is crucial because otherwise it would mean that the theory is always 
in a broken phase and therefore the FI-gauged symmetry is always ``Higgsed''. 
This can be avoided if we consider superspace higher derivative terms.

One type of terms we can consider are of the form 
\be
\label{HD-EX}
\begin{aligned}
\int d^4 \theta E & e^{2\xi V / 3 M_P^2} \left( \frac{\Phi}{M_P} \right)^{ \frac{\xi}{\tilde Q M_P^2} } W^2(V)  
\\ 
& \times |W^2(K+\Gamma)|^2  \left(\frac{\Phi e^{-\tilde Q V} \overline \Phi}{M_P^2} \right)^N + \text{c.c.}\,, 
\end{aligned}
\ee
where $|W^2|^2 = W^2 \overline W^2$ and $N$ is a positive integer large enough such that the limit $A \to 0$ is smooth. 
In addition, due to the excess of fermionic superfields, the specific higher order term will lead only to terms at least quadratic in the fermions, 
so it will not even influence the classical vacuum structure. 
In addition, as before, the fact that we have an excess of $\Phi$ superfields means that the global symmetry is broken, 
while the gauge invariance can be verified to be preserved. 
Note also that the formal global limit (i.e. the rigid supersymmetry limit) is well defined.

Let us note in passing that we have not discussed any anomaly cancellation here, since we assume that already in the theory with the global symmetry, or in any case, one can include the appropriate charges for the various fields such that gauge or gravitational anomalies are eliminated. The presence of the higher order terms does not influence the anomaly structure because we are not changing the measure of the path integral.

One may wonder what happens if we consider instead of $K$ and $W$ the generalized function $G = K + \log W + \log \overline W$ (which of course leads to the same component analysis). 
Clearly, such a setup requires that $W$ be nowhere vanishing. 
Instead, here we want to allow $W$ to vanish at least at one point in the moduli space such that the gauge vector is not always massive. 
This happens because $W$ is always charged under the FI-gauging as we see in \eqref{W-FI}. 
Therefore, working with $G$ is not an option if we want to keep the discussion general and focused on genuine FI-terms.

To close this general part, we have seen that there exist a couple of different terms that can break explicitly the global symmetry 
while maintaining the abelian gauge symmetry. 
One can check that there are also other types of terms that one can build with the same properties, increasing the number of fermions. 
From the mere existence of such terms, we can conclude that the existence of a global symmetry discussed in \cite{Komargodski:2009pc} is not enforced by the classical supergravity theory per se. 
Such a symmetry could of course appear accidentally in the low energy theory, 
but there is no reason to respect it within a supergravity model derived from some fully-fledged string theory compactification. 
There could be other conditions that do not allow the presence of a FI term or that enforce the existence of a global symmetry; 
however, the abelian gauge symmetry with the FI term is not enough to enforce such restrictions.

\section{A simple example}

In this section, we will discuss a simple setup where we can see the breaking of the global symmetry while the gauged symmetry is preserved. 
We will work first with the superpotential deformation because it is the easiest one to handle in component form, 
but it is by no means the unique deformation. 
Indeed, later on we will also analyse a superspace higher order term as well.

We simply consider a superfield $\Phi$ with 
\be
Q = 1 \ , \ R = 0 \,, 
\ee
such that the gauging takes the simple form 
\be
\label{KGex}
K+\Gamma = \overline \Phi e^{-V} \Phi + \xi V  \,, 
\ee
and we also have vanishing superpotential to start with, that is $W=0$. 
Notice that we have a global symmetry that acts as 
\be
\Phi \to \Phi \times \exp[ic_0] \,, 
\ee
which on the component level is 
\be
\label{GlobalU1}
A \to A \times \exp[ic_0] \quad , \quad \chi_\alpha \to \chi_\alpha \times \exp[ic_0] \qquad , \qquad F \to F \times \exp[ic_0] \, . 
\ee
Due to the compensating super-Weyl transformation, the actual symmetry that is gauged acts on the scalar component field in the following way: 
\be
\label{LocalU1A}
A \to A \times \exp[i \lambda] \, , 
\ee 
whereas for the fermion we find 
\be 
\label{LocalU1chi}
\chi_\alpha \to \chi_\alpha \times \exp\left[i \lambda + \frac{i}{2} \lambda \xi\right] . 
\ee
As a result, the gauged symmetry of the component form of the supergravity Lagrangian is not \eqref{GlobalU1}, 
but instead it is given by \eqref{LocalU1A} and \eqref{LocalU1chi}, which are manifestly different.

From our previous discussion we know that we can introduce a superpotential 
\be
\label{Wbreak}
W_{break} = b_0 \times (\Phi/M_P)^{- \frac{\xi}{M_P^2}} \,, 
\ee
which manifestly breaks the global symmetry \eqref{GlobalU1} but preserves the FI gauged U(1). 
Here we could also choose $b_0$ such that it vanishes when supergravity is decoupled or when $\xi$ goes to zero, 
e.g. $b_0 = \xi^2/M_P$, but this is not an essential requirement. 
Since there exists a superpotential with such a property, there exist infinite other higher derivative terms that can also do the same job. 
This shows that on the technical level supergravity allows for FI gaugings without global symmetries.

To see explicitly where the issue lies we can go to components and work with the superpotential \eqref{Wbreak} 
and the gauged K\"ahler potential \eqref{KGex}. 
In addition, we simply need fermionic terms to see the non-invariance, 
and due to the Grassmann nature of the fermions it is sufficient to look only at the quadratic terms. 
Since here we have chiral fermions which rotate to themselves under the global symmetry, we only need to look at the specific quadratic terms that include only the given chirality of these fermions. 
Therefore, we can simply look at the term (we set $M_P=1$ for simplicity) 
\be
\begin{aligned}
\label{Lchi^2}
e^{-1}{\cal L}_{\chi^2} = & - \frac{b_0}2 e^{|A|^2/2} \,  \chi^2  
\\
\times & \left[ 
\xi (\xi+1) A^{-\xi-2} 
- 2 \xi \overline A A^{-\xi-1} 
+ \overline A^2 A^{-\xi}  
\right] \,. 
\end{aligned}
\ee 
This term can be deduced following the standard rules of supergravity to go to component form \cite{Wess:1992cp}. 
Under the global rotation we find 
\be
\begin{aligned}
e^{-1}{\cal L}_{\chi^2}'  = & - \frac{b_0}2 e^{|A|^2/2} \,  \chi^2   e^{2 i c_0} \ \times e^{-(\xi+2) i c_0}  
\\ 
& \left[ \xi (\xi+1) A^{-\xi-2} 
- 2 \xi \overline A A^{-\xi-1} 
+ \overline A^2 A^{-\xi}  
\right] 
\\ 
 = & \, e^{-i \xi c_0} \times e^{-1}{\cal L}_{\chi^2}  \,. 
\end{aligned}
\ee
Therefore, the global rotation is not a symmetry of the full action. 

We can also see from the same term how the gauge invariance will work. 
Under the gauge transformation we have instead  
\be
\begin{aligned}
e^{-1}{\cal L}_{\chi^2}' = & - \frac{b_0}2 e^{|A|^2/2} \,  \chi^2   \exp\left[2 i \lambda + i \lambda \xi\right] \ \times e^{(-\xi-2) i \lambda} 
\\ 
& \left[ \xi (\xi+1) A^{-\xi-2} 
- 2 \xi \overline A A^{-\xi-1} 
+ \overline A^2 A^{-\xi}  
\right] 
\\ 
= & \, e^{-1}{\cal L}_{\chi^2}  \,. 
\end{aligned}
\ee
Thus, the term that breaks the global symmetry preserves the gauge invariance. 
Let us stress here that we do not presume that there will be some intricate effect that will generate such a superpotential. 
Instead, we are simply showing that on the classical supergravity level, 
there is no need for a global symmetry to exist, or in other words, a FI term does not inevitably lead to a global symmetry. 
The logic then here is similar to the typical approach we have for global symmetries in EFTs. 
Even if they are present at some order in our EFT, we know that they will certainly be broken by higher derivative/higher order terms once the system is coupled to gravity.

We can also expand the higher derivative term \eqref{HD-EX}. In the Wess-Zumino gauge, the terms of \eqref{HD-EX} that are quadratic in the spinor component of the chiral superfield $\Phi$ read
\begin{equation}
    \label{HDcomp}
     {\cal L}_{\text{HD},\chi^2} =
     {\cal L}_{A^{\xi + N}} + \text{c.c.}  \, , 
\end{equation}
where 
\begin{align}
    \label{Lxi+N}
    {\cal L}_{A^{\xi + N}} \equiv & \, \frac{1}{8}  e  A^{\xi + N} \overline{A}^N   \left[ \left( D_m A \right) \left( D^m A \right) \overline{\chi}^2 - \overline{F}^2 {\chi}^2 + 2 i \overline{F}  \chi \sigma^m \overline{\chi} D_m A \right] \nonumber \\
    & \times \left( 2 F_{np} F^{np}  + i \epsilon^{npqr} F_{np} F_{qr} - 4 D^2 \right) \nonumber \\
    & \times \big{[} 16 D_s A D^s A D_t  \overline{A} D^t \overline{A} + 16 i \left( A \overline{A} - \xi \right)  D_s A D_t \overline{A} F^{st} \nonumber \\
    & + 8  \epsilon^{stuw} \left( A \overline{A} - \xi \right) D_s A D_t \overline{A} F_{uw}   +16 D \left( A \overline{A} - \xi \right)  D_s A D^s \overline{A}  \\
    & - 32 F \overline{F}  D_s A D^s \overline{A} - 2 \left( A \overline{A} - \xi \right)^2 F_{st} F^{st} + i  \left( A \overline{A} - \xi \right)^2 \epsilon^{stuw} F_{st} F_{uw} \nonumber \\
    & + 4 \left( A \overline{A} - \xi \right)^2 D^2 - 16 \left( A \overline{A} - \xi \right) D F \overline{F} + 16 F^2 \overline{F}^2 \big{]} \, ,\nonumber 
\end{align}
with $D_m A = \partial_m A - \frac{i}{2} v_m A$. 
Here we have focused on the $\chi^2$ terms and ignored terms with the gaugino or the gravitino and all higher order fermionic terms. 
As we already discussed earlier, it is enough to focus on these terms to see the explicit breaking of the global symmetry.

Indeed, the contribution \eqref{HDcomp} to the higher derivative term is not invariant under the global transformation \eqref{GlobalU1}, since it transforms as 
\begin{equation}
    \label{HDglobalU1}
     {\cal L}_{\text{HD},\chi^2} \to e^{i \xi c_0} {\cal L}_{A^{\xi + N}} + \text{c.c.} \,. 
\end{equation} 
On the other hand, under an abelian gauge transformation that preserves the Wess-Zumino gauge followed by a compensating super-Weyl transformation, parametrized by the chiral superfield \eqref{R-G}, the auxiliary complex scalar component field of $\Phi$ transforms as 
\begin{equation}
    \label{LocalU1F}
    F \to F \times \exp[i(1+\xi)\lambda]  \,, 
\end{equation}
the vector field $v_m$ transforms as
\begin{equation}
    \label{LocalU1v}
    v_m \to v_m + 2 \partial_m \lambda \, , 
\end{equation}
while the auxiliary real scalar $D = - \frac{1}{2} {\cal D}^{\alpha} W_{\alpha} (V)|$ and the vielbein are invariant. Using the transformation rules \eqref{LocalU1A}, \eqref{LocalU1chi}, \eqref{LocalU1F} and \eqref{LocalU1v} it is straightforward to verify that the part \eqref{HDcomp} of the higher derivative term \eqref{HD-EX} is gauge invariant.

One can also verify the properties of such terms when we integrate out the auxiliary fields. 
We can simply evaluate them on the background that is generated 
by the 2-derivative supergravity bosonic sector of \eqref{stand} with $K + \Gamma$ given by \eqref{KGex} and $W=0$. This happens because as far as the 2-fermion terms are concerned, the auxiliary fields will only contribute via their bosonic values. 
If we kept higher order terms in the fermions in the on-shell Lagrangian, then we would also need  the fermionic contributions to the on-shell values of  the auxiliary fields. 
Focusing therefore on the bosonic part of the auxiliaries, we find 
\be
\label{F,D}
F = 0 \quad , \quad D =\frac{1}{2} e^{-\frac{1}{3} A \overline{A}} \left( A \overline{A} - \xi \right)  .
\ee
Inserting these back into the higher order expression and performing the required Weyl-rescaling and fermion redefinition of \cite{Wess:1992cp} we get 
\begin{align}
\label{HDcompresc}
e^{-1}{\cal L}_{\text{HD},\chi^2} = & \, \frac{1}{8} e^{- \frac{7}{6} A \overline{A}} A^{\xi + N} \overline{A}^N   \left( D_m A \right) \left( D^m A \right) \overline{\chi}^2 \nonumber \\ 
& \times \left[ 2 F_{np} F^{np}  + i \epsilon^{npqr} F_{np} F_{qr} - \left( A \overline{A} - \xi \right)^2 \right] \nonumber \\
    & \times \Big{[} 16 D_s A D^s A D_t  \overline{A} D^t \overline{A} + 16 i \left( A \overline{A} - \xi \right)  D_s A D_t \overline{A} F^{st} \nonumber \\
    & + 8  \epsilon^{stuw} \left( A \overline{A} - \xi \right) D_s A D_t \overline{A} F_{uw}   +8  \left( A \overline{A} - \xi \right)^2  D_s A D^s \overline{A}  \\
    & - 2 \left( A \overline{A} - \xi \right)^2 F_{st} F^{st} + i  \left( A \overline{A} - \xi \right)^2 \epsilon^{stuw} F_{st} F_{uw} \nonumber \\
    & +  \left( A \overline{A} - \xi \right)^4  \Big{]} + \text{c.c.} \nonumber \, .  
\end{align}
One can once again verify that the global symmetry is broken, while gauge invariance is maintained. In addition, notice that crucially this term has manifestly a smooth $A \to 0$ limit.

\section{The Komargodski--Seiberg theorem}

 In Ref.\cite{Komargodski:2009pc}, Komargodski and Seiberg gave a generic proof for the inconsistency of supergravity theories with FI terms. According to their line of reasoning,  such  theories  always preserve a global symmetry, and therefore are  in clash with quantum gravity. Let us recall their argument and see how it is evaded in our case. 

Let us consider  a supergravity theory with a FI term, which is  described by a Lagrangian $\mathcal{L}_\xi$. The theory also has a $U(1)_{\rm FI}^{\xi}$ gauge symmetry, which is generated by  $\delta_\xi$ such that 
\be
\delta_\xi \mathcal{L}_\xi=0. \label{xiL}
\ee
Let us further assume that  both $\mathcal{L}_\xi$ and $\delta_\xi$ have a perturbative expansion in the FI parameter $\xi$ of the form
\be
\mathcal{L}_\xi=\mathcal{L}_0+\xi\mathcal{L}_1+\cdots\, , \qquad
\delta_\xi=\delta_0+\xi \delta_1+\cdots\,  .
\ee
Then, the invariance of the theory, as described by (\ref{xiL}), leads to the conditions
\begin{align}
&\delta_0\mathcal{L}_0=0 \, , \label{ll1}\\
&\delta_1\mathcal{L}_0+\delta_0 \mathcal{L}_1=0 \, . \label{ll2}
\end{align}
Acting with $\delta_0$ on (\ref{ll2}),  using (\ref{ll1}) and the fact that for the linear transformation $\delta_\xi$  the order  is irrelevant, i.e. $[\delta_0,\delta_1]=0$, we get that 
\be
\delta_0^2 \mathcal{L}_1=0 \,.
\ee
Now $\delta_0$ acts as a rotation 
% $\delta_0 \mathcal{L}_1=\lambda  \mathcal{L}_1$, we get 
so that $\delta_0\mathcal{L}_1=0$, and therefore,   
\be
\delta_0\mathcal{L}_0=\delta_0\mathcal{L}_1=0 \, .
\ee
We can continue the above process for all $\mathcal{L}_i$ so that finally
\be
\delta_0 \mathcal{L}_\xi=0 \,. 
\ee
In other words, if there is a perturbative expansion of the Lagrangian of the theory with respect to $\xi$, then on top of the gauge symmetry generated by $\delta_\xi$, there is also a global symmetry generated by $\delta_0$. Therefore,  all supergravity theories must have vanishing FI-terms, since the existence of the latter leads always to a global symmetry, which is inconsistent with quantum gravity.  

In our case, the theorem is evaded since 
$\delta_0$ does not act as a rotation. Indeed, let us assume that $\mathcal{L}_1$ contains a logarithm of the fields: 
\be
\mathcal{L}_1\supset \log A \, .
\ee
Clearly, after a global  rotation $e^{ia} A$ of the $A$ field, generated by $\delta_0$, $\mathcal{L}_1$ changes as 
\be
\delta_0 \mathcal{L}_1\sim i a,
\ee
but still $\delta_0^2 \mathcal{L}_1=0.$
Therefore, $\delta_0$ does not generate a global symmetry. This is the situation we have here as for example the superpotential (\ref{W-break}) leads to 
\be
\mathcal{L}_1\supset -\frac{b_0}{Q M_P^2}\log A. 
\ee
In this case, $\mathcal{L}_1$  gives 
\be
\delta_0 \mathcal{L}_1 \sim -\frac{b_0}{Q M_P^2}i \alpha,
\ee
and therefore, the supergravity with the FI terms does not lead to any inconsistency, at least due to a global symmetry, which is violated in our case.

In other words, 
even if the alleged global symmetry acts as a rotation on the fields independently, 
such a property may not hold for the $\xi$-expanded functions of the fields that enter the Lagrangian. In particular, this is the case for   $A^\xi = 1 + \xi \log A + {\cal O}(\xi^2)$. 
Clearly, the order-$\xi$ part proportional to $\log A$, does not change like a rotation under the alleged global symmetry, but instead it shifts, and thus  the KS argument does not apply.

To be more specific, one can work out \eqref{Lchi^2} along the lines above.
%explicitly as we have done in the previous sections that indeed there always exists a series of terms that can explicitly break the continuous global symmetry, but also apply our discussion above to specific terms to see the loophole. 
Using the Taylor series expansion 
\begin{equation}
    \label{A^(-xi)}
    A^{- \xi} = e^{- \xi \log A} = 1 - \xi \log A + \mathcal{O}(\xi^2)
\end{equation}
we can expand the term \eqref{Lchi^2} in the parameter $\xi$ as 
\begin{equation}
    \label{Lxichi}
    {\cal L}_{\chi^2} = {\cal L}_0 + \xi {\cal L}_1 + \mathcal{O} (\xi^2) \, , 
\end{equation}
where 
\begin{align}
    \label{L0}
    {\cal L}_0 & = - \frac{1}{2} b_0 e e^{|A|^2/2} \overline{A}^2 \chi^2  , \\
    \label{L1}
    {\cal L}_1 & = - \frac{1}{2} b_0 e e^{|A|^2/2} ( A^{-2} - 2 \overline{A} A^{-1} - \overline{A}^2 \log A ) \chi^2 .
\end{align}
Furthermore, an infinitesimal abelian gauge transformation with parameter $\lambda(x)$ acts on the fields $A$ and $\chi_{\alpha}$ as 
\begin{align}
    \label{dxiA}
    \delta_{\xi} A & = i \lambda A \equiv \delta_0 A \, , \\
    \delta_{\xi} \chi_{\alpha} & = i \left( 1  + \frac{\xi}{2} \right) \lambda \chi_{\alpha} \equiv \delta_0 \chi_\alpha + \xi \delta_1 \chi_\alpha \, , 
\end{align}
where 
\begin{equation}
    \label{d0d1chi}
    \delta_0 \chi_{\alpha} = 2 \delta_1 \chi_{\alpha} = i \lambda \chi_{\alpha} \, . 
\end{equation}
Using \eqref{dxiA} and \eqref{d0d1chi} it is straightforward to show that 
\begin{equation}
    \label{d0L0}
    \delta_0 {\cal L}_0 = 0 
\end{equation}
and 
\begin{equation}
    \label{d0L1}
    \delta_0 {\cal L}_1 = - \, \delta_1 {\cal L}_0 = \frac{i}{2} b_0 \lambda e e^{|A|^2/2} \overline{A}^2 \chi^2 . 
\end{equation}
From the form of $\delta_0 {\cal L}_1$ we see that acting with one more $\delta_0$ we find 
\begin{equation}
    \label{d0^2L1}
    \delta_0^2 {\cal L}_1 = 0 \, ,
\end{equation}
even though $\delta_0 {\cal L}_1$ is non-vanishing. 
Therefore, this case appropriately illustrates how the KS argument is surpassed. 
Clearly, the terms of higher derivative nature also work in the same way, and therefore, we do not analyze them further here.

\section{Discussion}

We have seen that in principle, the presence of a FI term will not lead to the compulsory preservation of a global symmetry. 
This happens because the gauge invariance works independently of the global symmetry, and therefore there can exist in principle higher order terms (possibly Planck-suppressed) that break the latter while preserving the former.

However, 
even if the global symmetry is eliminated, there is still the question of the compactness of the abelian FI gauge invariance. 
Clearly, as long as $\xi / M_P^2$ is some small perturbative number, it is generically impossible to have a compact abelian gauge invariance. 
There is of course the trivial case $Q_i=0=R_i$, 
which also eliminates the need for the extra breaking of the global symmetry. 
There is nevertheless one instance where compactness could be non-trivially preserved, and that is if one arranges for all the shifted bosonic charges to vanish. 
Indeed, setting $\tilde Q_i=0$ in \eqref{tildeQ}, requires to have special theories such that $R_i = -2 M_P^2 Q_i / \xi$, 
but leaves only the fermions (chiralini, gaugini, gravitino) charged with charges $\pm \xi / 2 M_P^2$ thus the group becomes compact. 
The latter setup of course will have a pathological rigid limit.

Our work here was primarily focused on a setup where $\xi$ is not quantized in terms of $M_P$. 
If instead one turns to systems with a quantized FI parameter then it is already known that we need $\xi = - 2 n M_P^2$ 
to have a compact group, for $n \in \mathbb{Z}$ \cite{Seiberg:2010qd,Distler:2010zg}. 
In this case, the superpotential from \eqref{W-break} becomes $W_{break} \sim \Phi^{\frac{2 n}{Q-n R}}$, 
and one finds similar expressions for the higher derivative terms. 
Here in principle one will not face some extra restrictions; 
however, 
there can be special choices for the R-charges and the Q-charges which lead to residual global discrete symmetries; 
such choices will belong to the swampland. 
Indeed, 
if we have $n=1$ with $R=0$ and $Q=1$ then we see that $W_{break} \sim \Phi^2$ therefore the system has a $Z_2$ symmetry. 
Notably, this $Z_2$ is also preserved if we include higher derivative terms such as \eqref{HD-EX}. 
If instead $n=1$ with $R=-1$ and $Q=1$, then we have $W_{break} \sim \Phi$ and therefore there is no residual discrete symmetry.

To recap, 
we have seen that both for perturbative and for quantized $\xi$ we can anyhow break the continuous global symmetry of \cite{Komargodski:2009pc}. 
In the case of perturbative $\xi$, the group is non-compact, which remains an issue as far as the coupling to quantum gravity is concerned \cite{Banks:2010zn}; 
we do not have a way at this point to resolve this problem (while preserving supersymmetry). 
Since the existence of the FI term in the rigid limit requires $\xi$ to be independent of $M_P$, one is tempted to say 
that a FI term in the rigid limit cannot arise from a genuine FI term in the supergravity theory (unless the latter is in the swampland). 
If instead $\xi$ is quantized in terms of $2 M_P^2$, we have seen that there are always ways to break the continuous global symmetry. 
One then has to impose some simple limitations on the spectrum such that the residual discrete symmetry is also broken. 
Of course having a quantized $\xi$ means the rigid limit will generically be pathological.

\section*{Acknowledgments} 

We thank Dmitri Sorokin and Antoine Van Proeyen for feedback and discussions. The research work was supported by the Hellenic Foundation for Research and Innovation (HFRI) under the 3rd Call for HFRI PhD Fellowships (Fellowship Number: 6554).
\begin{center}
\includegraphics[scale=0.35]{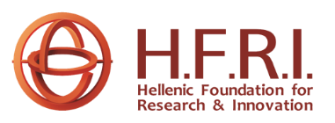}
\end{center}

\end{document}